\documentclass[conference]{IEEEtran}
\ifCLASSINFOpdf
\else
\fi

\usepackage{cite}
\usepackage{mathtools}
\usepackage{amsmath}
\usepackage{graphicx}
\usepackage{amsmath}
\usepackage{amssymb}
\usepackage[noend]{algpseudocode}
\usepackage{algorithmicx,float}
\usepackage{verbatim}
\usepackage{lipsum,amsmath,multicol}
\usepackage{color,soul}
\usepackage{qtree}
\usepackage{xcolor}
\usepackage{amsthm}
\usepackage{enumitem}
\usepackage{algorithm,algpseudocode,float}
\usepackage{pifont}
\usepackage{subcaption}
\usepackage{caption}
\usepackage{hyperref}

\begin{document}
%
\title{Joint Optimization of QoE and Fairness Through Network Assisted Adaptive Mobile Video Streaming}

\author{\IEEEauthorblockN{Abbas Mehrabi, Matti Siekkinen, and Antti Yl\"a-J\"{a}\"{a}ski}
\IEEEauthorblockA{Department of Computer Science, Aalto University\\
P.O.Box 15400, FI-00076, Espoo, Finland\\
Emails: \{abbas.mehrabidavoodabadi, matti.siekkinen, antti.yla-jaaski\}@aalto.fi}
}


%


\maketitle

\begin{abstract}
MPEG has recently proposed Server and Network Assisted Dynamic Adaptive Streaming over HTTP (SAND-DASH) for video streaming over the Internet. In contrast to the purely client-based video streaming in which each client makes its own decision to adjust its bitrate, SAND-DASH enables a group of simultaneous clients to select their bitrates in a coordinated fashion in order to improve resource utilization and quality of experience. In this paper, we study the performance of such an adaptation strategy compared to the traditional approach with large number of clients having mobile Internet access. We propose a multi-servers multi-coordinators (MSs-MCs) framework to model groups of remote clients accessing video content replicated to spatially distributed edge servers. We then formulate an optimization problem to maximize jointly the QoE of individual clients, proportional fairness in allocating the limited resources of base stations as well as balancing the utilized resources among multiple serves. We then present an efficient heuristic-based solution to the problem and perform simulations in order to explore parameter space of the scheme as well as to compare the performance to purely client-based DASH. 
\end{abstract}
 

%
\IEEEpeerreviewmaketitle

\section{Introduction}

According to statistics, the majority of Internet traffic is video, such as Netflix, YouTube, or other streaming applications~\cite{Sandvine2012}, \cite{Sandvine2013}. The network conditions, such as high fluctuation in the available bandwidth when multiple clients simultaneously compete for the shared bottleneck link, can significantly affect the users' quality of experience (QoE) in mobile video streaming applications \cite{Yao2011}, \cite{Riiser2012}. Mobile and wireless access further complicate the situation. 
In order to avoid playback interruption and rebuffering events due to changes in available bandwidth during a streaming session, most media players nowadays use adaptive streaming, such as the non-standard HTTP Live Streaming (HLS) or protocols based on the dynamic adaptive video streaming (DASH) standard. In adaptive streaming, the whole video is divided into chunks and encoded with different qualities on the server~\cite{Seufert2015}. The client adapts dynamically to bandwidth fluctuations by downloading appropriate bitrate chunks and therefore improving the QoE of end users~\cite{Sun2016}.   

Numerous research efforts, both theoretical and experimental, have been carried out in recent years on designing efficient adaptation mechanisms for mobile video streaming~\cite{Jiang2012,Huang2014,Wang2016}. 
Almost all of them focus on improving the client-side adaptation strategy. 
However, recent innovations in mobile network architectures and cloud computing, namely Mobile Edge Computing (MEC)\cite{Tran2017} and Fog Computing \cite{FogComputing}, provide an opportunity to further optimize the content delivery and adaptation through in-network and edge computing mechanisms. The recently proposed Server and network assisted DASH (SAND-DASH) standard specifies mechanisms and message types so that clients, network, and servers can exchange information and collaborate in video quality adaptation in order to improve QoE and fairness~\cite{Thomas2016}. However, the standard specifies no adaptation logic, which is left open for innovation on purpose, and some preliminary work has already emerged studying different network assisted adaptation mechanisms~\cite{Cofano2016}. 

Our overarching goal is to understand how much, in which way, and at which cost (esp. computational complexity) the QoE and fairness between mobile video streaming clients could be improved through network and server assisted adaptation mechanisms compared to pure client-based mechanisms. To this end, we present in this paper a general framework for studying network-assisted quality adaptation of large number of mobile DASH clients streaming replicated video content from mobile edge servers. We first formulate an optimization problem for jointly maximizing the QoE of individual clients, the proportional fair (PF) resource allocation at the base stations as well as balancing the utilized resources among multiple servers. Then, we design an efficient solution to this problem and compare its performance to purely client-based adaptation heuristics, namely rate-based and buffer-based adaptation, and examine the parameter space of the proposed solution using simulations.

The rest of the paper is organized as follows: We discuss related work 
in Section \ref{sec:rw} and describe the proposed framework and its components in Section \ref{sec:framework}.  
The optimization problem is laid out in Section \ref{sec:problem} and its solution is detailed in Section \ref{sec:solution}. 
We present simulation-based evaluation in Section \ref{sec:eval} before concluding the paper and pointing out avenues for future work.

\section{Related Work}
\label{sec:rw}

Several approaches for adaptive video streaming have been proposed during the past years \cite{Huang2014}, \cite{Mangla2016}, \cite{Sun2016}, \cite{Bentaleb2016}, \cite{Jiang2012}, \cite{Wang2016}. Seufert et al.~\cite{Seufert2015} provide a comprehensive study on video quality adaptation and the major QoE related factors that both client and network need to consider. While the first generation of adaptive streaming protocols, such as HLS, select video bitrate merely based on measured download rate of previous chunks, Huang et al.~\cite{Huang2014} proposed an adaptation strategy that selects a video bitrate purely based on the current playback buffer occupancy level. Spetiri et al.~\cite{Spiteri2016} designed an online bitrate adaptation algorithm, BOLA, that is also based on buffer occupancy level only and prove its performance guarantee.  
Since purely buffer-based adaptation mechanisms may be sub-optimal especially under high throughput fluctuation, some techniques combine the buffer occupancy level with throughput prediction~\cite{Sun2016,Jiang2012,Wang2016}. 

Several recent papers have investigated quality adaptation considering multiple clients associated with either single or multiple video servers~\cite{Petrangeli2015,Bethanabhotla2015,Bouten2014}. Petrangeli et al.~\cite{Petrangeli2015} examined the fair bandwidth utilization when multiple clients compete on shared bottleneck link, but their proposed objective function and the adaptation heuristic fail to capture the 
trade-off between client-perceived QoE and fair network resource utilization. The objective of bitrate selection by Bethanabhotla et al.~\cite{Bethanabhotla2015} is to maximize video qualities subject to the stability of servers' queues without considering 
the instantaneous and dynamic nature of bandwidth fluctuation, which according to \cite{Yao2011} has direct impact on the QoE. 
Bouten et al.~\cite{Bouten2014} propose in-network optimization of clients' bitrates according to the available bandwidth on multiple bottleneck links. However, in this work, only one bitrate is allocated to each client and also each client is associated to a specified server which is known in advance. Furthermore, the proposed objective function neglects the initial buffering delay and the number of stalling events which have significant impact on the QoE of individual clients \cite{Seufert2015}.  

Concerning fair resource allocation in cellular networks, schedulers usually aim for proportional fairness (PF) when allocating radio resources to multiple competing clients in order to balance cell throughput with fairness. Chen et al.~\cite{Chen2013} propose a scheduling framework called AVIS that strives for proportional fairness while controlling the number of bitrate switching in scheduling of multiple simultaneous clients. However, the competition for accessing the available resources is considered on the shared bottleneck of only one base station and also this work mostly focuses on the fair resource scheduling without considering QoE-related parameters such as the initial buffering delay or the buffer stalling.

The recently proposed Server and network assisted DASH (SAND-DASH) standard specifies means for clients, network elements, and servers to exchange information in order to optimize video delivery and quality adaptation. It does not specify any adaptation logic but some work already exists on trying to understand the effectiveness of this approach~\cite{Li2016,Cofano2016,Thomas2016,Bentaleb2016}. Our work contributes to these efforts and, to the best of our knowledge, is the first one to try to quantify the benefits of network assisted quality adaptation in mobile video streaming with edge caching.

\section{Network-Assisted Mobile Video Streaming}
\label{sec:framework}

\subsection{Multi-Servers Multi-Coordinators (MSs-MCs) Framework}
\label{sec:fw}

\begin{figure}[t]
\begin{center}
\includegraphics[trim = 0in 0in 0in 0in, clip=true, width=3.6in, height=2.3in]{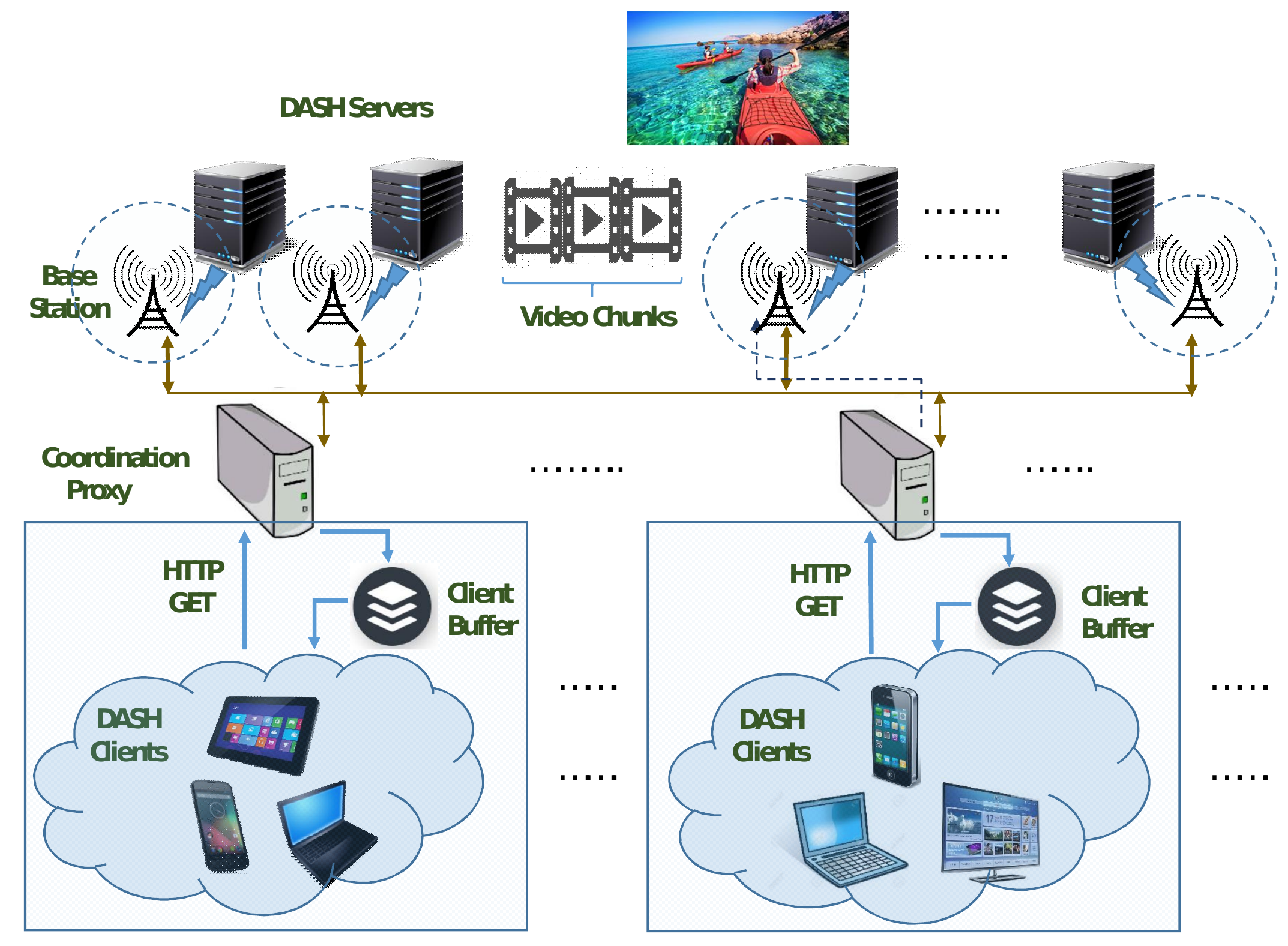}
\caption{Multi-servers multi-coordinators (MSs-MCs) framework for dynamic  
adaptive video streaming at large scale.}
\label{fig:framework}
\vspace{-2em}
\end{center}
\end{figure}

Fig. \ref{fig:framework} illustrates the proposed multi-servers multi-coordinators (MSs-MCs) framework for network assisted adaptive video streaming. The DASH servers at the top store the replicated videos. We assume that a discrete set of $M$ videos denoted by $V=\{v_{1},v_{2},...,v_{M}\}$ are divided into multiple chunks with fixed size $C$ (in Seconds) and replicated on $K$ mobile edge servers each of which is associated to a base station. The base station allocates available radio resources in a proportionally fair manner to clients~\cite{Chen2013}. Each server stores video chunks with multiple bitrate resolutions such that $Q_{k}$ 
denotes the discrete set of bitrate resolutions for every video chunk which is offered by server $k$. We partition the potentially large mobile network and its clients into subnets and groups. Considering $N$ groups (subnets) $G_{1},G_{2},...,G_{N}$ of DASH clients distributed over a potentially large geographical area, the clients join the network so that $|G_{j}|, 1 \leq j \leq N$ denotes the number of currently active clients in subnet $j$. From the practical point of view, the clients which are in close vicinity are managed as one group and their information (arrival/departure times, physical location, buffer occupancy) is exchanged with the central scheduler located on the cloud through the local coordination proxy on the edge of the network. Following the discrete time slotted DASH scheduling \cite{Bouten2014} and with total number of $|T|$ time slots, at each time slot $1 \leq t \leq |T|$, the data transmission between the base station associated with video server $k$ and different clients goes through a shared bottleneck link with capacity of $W_{k}^{(t)}$. Please note that $W_{k}^{(t)}$ refers to the available resource blocks i.e. the number of subcarriers in the frequency domain, at time slot $t$ on base station $k$. We also note that the clients are assumed to be stationary throughout this work while we consider the clients' mobility and the impact of handover as one of our interesting future works. 

Let $A_{ij}$ and $D_{ij}$ denote the arrival and the departure times, respectively, of client $i$ belong to network $j$ which correspond to the time that client sends its request for first chunk and the time that it either abandons the streaming session or finishes downloading the last chunk. In the ideal case when no stalling happens during the session and with negligible network delay, the quantity $|D_{ij}-A_{ij}|$ is obviously equal to the watching duration of the video requested by client $i \in G_{j}$ and consequently $|D_{ij}-A_{ij}|/C$ is the number of streaming chunks of the video. 
The media player of each client $i \in G_{j}$ maintains a playback buffer for which the client determines a fixed target filling level denoted by $B_{ij}^{max}$ (in Kb). $B_{ij}^{(t)} \leq B_{ij}^{max}$ represents the level of data in the client's buffer at time slot $t$. 
The coordination proxy does the client to server mapping based on client information (buffer occupancy, radio link conditions), QoE metrics considered, proportional fair bitrate allocation at the base stations as well as load balancing between servers. 
For the client to server mapping, we define a binary variable $x_{ijk}^{(t)}$ such that $x_{ijk}^{(t)}=1$ if client $i \in G_{j}$ is allocated to server $k$ for downloading the current chunk at time slot $t$ and $x_{ijk}^{(t)}=0$, otherwise. Furthermore, the integer decision variable $r_{ijk}^{(p)} \in Q_{k}$ denotes the allocated bitrate for chunk index $p$ which is downloaded by client $i \in G_{j}$ from server $k$. 
 
Before we formulate the optimization problem in Section \ref{sec:problem}, we discuss next the different optimization criteria related to QoE, fairness, and server load balancing. 

\subsection{Quality of Experience}
\label{sec:bg}



A recent comprehensive study on QoE in dynamic adaptive video streaming \cite{Seufert2015} shows that four major factors can significantly 
\begin{table}[t]
\renewcommand{\arraystretch}{1.3}
\caption{Description of parameters in MSs-MCs framework.}
\label{table1}
\centering
\begin{scriptsize}
\begin{tabular}{|p{2cm}|l|}
\hline
Notation & \multicolumn{1}{|c|}{Description} \\
\hline\hline
$V=$ & \multicolumn{1}{p{6cm}|}{Set of $M$ available videos}  \\
 $\{v_{1},v_{2},...,v_{M}\}$  &   \\
\hline
$C$ & \multicolumn{1}{p{6cm}|}{Constant size of each video chunk (in Seconds)}  \\
\hline 
$K$, $N$, $S$, $Q_{k}$   &  \multicolumn{1}{p{6cm}|}{Number of servers, groups, clients and the discrete set of video bitrates offered by server $k$, respectively}    \\
\hline
$G_{j}$, $|G_{j}|$   & \multicolumn{1}{p{6cm}|}{Group (subnet) $j, 1 \leq j \leq N$ and its size} \\
\hline
$|T|$   & \multicolumn{1}{p{6cm}|}{Total number of scheduling time slots} \\
\hline
$W_{k}^{(t)}$  &   \multicolumn{1}{p{6cm}|}{Available resource blocks at base station $k$ in time slot $t$}  \\
\hline
\hline
$A_{ij}$, $D_{ij}$   & \multicolumn{1}{p{6cm}|}{Arrival and departure times of client $i \in G_{j}$} \\
\hline
$B_{ij}^{max}$   &  \multicolumn{1}{p{6cm}|}{Maximum buffer filling level of client $i \in G_{j}$}  \\
\hline
$B_{ij}^{(t)}$   &  \multicolumn{1}{p{6cm}|}{Buffer level of client $i \in G_{j}$ at time slot $t$}  \\
\hline
$AQ_{ij}$   &  \multicolumn{1}{p{6cm}|}{Average video quality for client $i \in G_{j}$}  \\
\hline
$Delay_{ij}$   &  \multicolumn{1}{p{6cm}|}{Initial buffer delay for client $i \in G_{j}$}  \\
\hline
$E_{ij}$    &  \multicolumn{1}{p{6cm}|}{Accumulated bitrate switching for client $i \in G_{j}$} \\
\hline
$d_{ijk}^{(t)}$  &  \multicolumn{1}{p{6cm}|}{Physical distance of client $i \in G_{j}$ from base station $k$ at time slot $t$} \\
\hline
$Thr_{ijk}^{(t)}, \hat{Thr}_{ijk}^{(t)}$   &  \multicolumn{1}{p{6cm}|}{Theoretical and effective received data throughput by client $i \in G_{j}$ from base station $k$ at time slot $t$} \\
\hline
\hline
$F_{ij}$   &  \multicolumn{1}{p{6cm}|}{Proportional fairness contribution by client $i \in G_{j}$} \\
\hline
$U_{ijk}$  &  \multicolumn{1}{p{6cm}|}{Percentage of utilized resources by client $i \in G_{j}$ on server $k$}   \\
\hline
$\bar{U_{ij}}$    &  \multicolumn{1}{p{6cm}|}{Average utilized resources by client $i \in G_{j}$}  \\
\hline
\hline
$P_{max}$   &  \multicolumn{1}{p{6cm}|}{Maximum transmission power of each base station}  \\
\hline
$\alpha$    &  \multicolumn{1}{p{6cm}|}{Path loss exponent parameter (normally between 2 and 5)}  \\ 
\hline
\hline
$\rho$, $\omega$, $\gamma$   &  \multicolumn{1}{p{6cm}|}{Adjustable weighting parameters for average quality, initial delay and bitrate switching, respectively}  \\
\hline
$\beta$, $\theta$, $\mu$    &  \multicolumn{1}{p{6cm}|}{Adjustable weighting parameters for QoE, proportional fairness and load balancing, respectively}   \\
\hline
\hline
$x_{ijk}^{(t)}$   &   \multicolumn{1}{p{6cm}|}{Binary allocation of client $i \in G_{j}$ to server $k$ at slot $t$}  \\
\hline
$r_{ijk}^{p} \in Q_{k}$    &   \multicolumn{1}{p{6cm}|}{Discrete allocated bitrate for chunk index $p$ of client $i \in G_{j}$ by server $k$} \\
\hline    
\end{tabular}
\end{scriptsize}
\end{table}
affect the quality of experience perceived by DASH clients: \emph{video quality}, \emph{startup delay}, \emph{stalling ratio} and \emph{quality switching}. 



\textbf{Video quality} is dependent on the video bitrate but the relationship is not necessarily linear~\cite{Bentaleb2016}. 
There is a trade-off between video quality and stalling: Streaming high quality video increases the probability of experiencing a stall event because the download throughput has a higher chance to drop below the video bitrate due to low bandwidth available on the bottleneck link. Streaming at low quality reduces the possibility of stalling but also significantly degrades the client's quality of experience. On the other hand, bitrate does not directly express the video quality and we need a function $q=f(r)$ that maps a bitrate $r$ to a quality $q$. In Section \ref{sec:eval}, we use the $log$ function adopted from \cite{Bentaleb2016} as the Structure Similarity (SSIM) index for the mapping function $f$.      

\textbf{Startup delay} refers to the time duration which is needed to reach the target buffer filling level of the client upon its arrival. It corresponds to the waiting time of client from click to start of the playback. According to~\cite{Hossfeld2012}, the startup delay has a clearly smaller impact on the dissatisfaction of a viewer than stall events.

\textbf{Stalling ratio} is the the amount of time spent so that video playback is stalled divided by the total duration of the session. Stall events occur when playback buffer empties caused by too low download throughput compared to the video bitrate. Avoiding stall events is critically important because of their prominent role in determining QoE. Therefore, we design the optimization problem with such constraints that stall events are avoided whenever possible, i.e. whenever the total amount of resources suffices to support lowest available video bitrates for all clients. 

\begin{figure}[t]
\begin{center}
\includegraphics[trim = 0in 0in 0in 0in, clip=true, width=2.6in, height=2.1in]{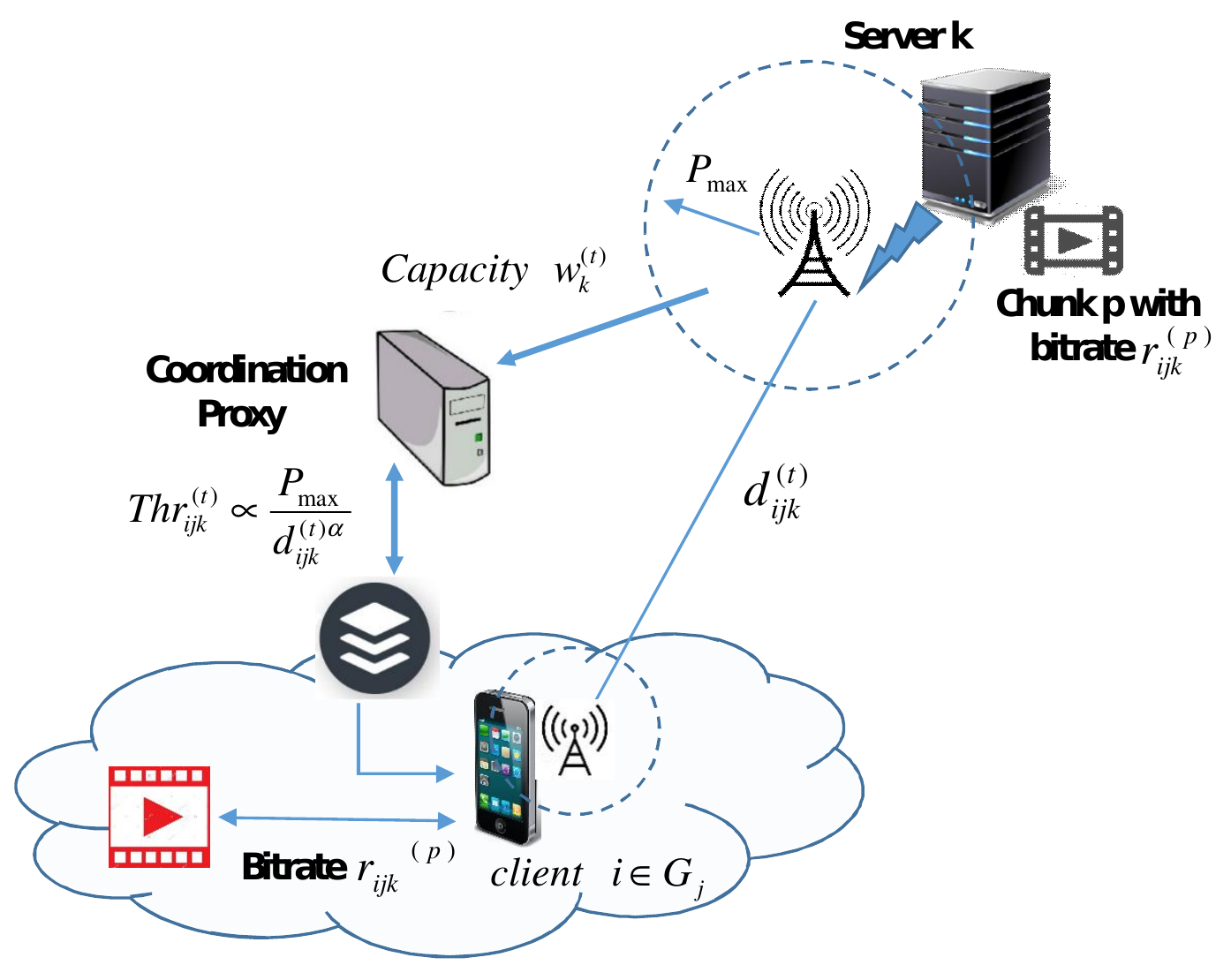}
\caption{Theoretical received data throughput at client's buffer.}
\label{fig:throughputcomputation}
\vspace{-2em}
\end{center}
\end{figure}

Frequent \textbf{quality switching} is also considered harmful for QoE~\cite{Seufert2015}. To express it as QoE metric, we consider the difference between the quality level of consecutive chunks of the video downloaded by the client as the metric. 

\subsection{Proportional Fairness}

In cellular networks, such as LTE, the base stations usually schedule radio resources to multiple competing clients at each time slot according to a proportional fairness (PF) policy~\cite{Chen2013}. Specifically, the amount of resources allocated to a client is proportional to its link quality (data rate). It is remarked that in a different way from \mbox{\cite{Colonnese2017}} in which the optimized bandwidth allocation is investigated on the base station side, we consider the QoE-aware optimal bitrate allocation taking into account the PF resource allocation policy by the base stations.  

\subsection{Sever Load Balancing}

We also consider load balancing between the servers as a criteria for optimization in order to avoid situations where a particular server's computational capacity becomes a bottleneck. We consider at each time instant the percentage of the resources used out of the total available resources on each server as the metric for resource utilization for that server \cite{Chen2013}.

\section{Joint Optimization Problem}
\label{sec:problem}

In formulating the optimization problem, we consider the four QoE metrics, fairness, and load balancing discussed in the previous section. Please, note that the optimization framework is independent of the way the metrics are computed. In order to balance the impact of these main factors, we define three weighting parameters $\beta$, $\theta$ and $\mu$ that control the relative importance of resulting QoE metrics, proportional fairness in bitrate allocation, and server load balancing. We further define three adjustable weighting parameters $0 \leq \rho, \omega, \gamma \leq 1$ to control the individual QoE metrics, namely video quality, initial playback delay, and the accumulated quality switching. In addition, we include constraint in the optimization problem in order to avoid stall whenever possible.   

The problem formulation follows a discrete time slotted scheduling operation with fixed time duration $\Delta t=1 Seconds$ of each time slot. We define the problem as a utility maximization problem over all clients $i \in G_{j}$ using the integer non-linear programming (INLP) formulation in \eqref{eq:opt}. The variables $x_{ijk}^{(t)}$ and $r_{ijk}^{(p)}$ are the binary and integer decision variables, respectively, while the values of other parameters are known in advance. We next explain how we obtain each individual parameter and constraint.  
\begin{align}
\label{eq:opt}
\small
Maximize \hspace{3mm} Utility(i,j)= & \beta (\rho AQ_{ij} - \omega Delay_{ij} - \gamma E_{ij})   \nonumber \\
\indent \hspace{5mm} & +  \theta \cdot  PF_{ij} - \mu \cdot SD_{ij}
\end{align}
Subject to:
\begin{align}
\small
&\sum_{k=1}^{K} x_{ijk}^{(t)}=1, \hspace{5mm} \forall A_{ij} \leq t \leq D_{ij}  \label{eq:constr1}\\
&\sum_{t^{\prime}=\lfloor \frac{t}{C} \rfloor \cdot C+1}^{\lceil \frac{t}{C} \rceil \cdot C} x_{ijk}^{(t^{\prime})}=\{0,C\}, \hspace{2mm} \forall 1 \leq k \leq K, A_{ij} \leq t \leq  D_{ij}  \label{eq:constr2}\\
&\sum_{t=A_{ij}}^{A_{ij}+Delay_{ij}} \sum_{k=1}^{K} x_{ijk}^{(t)} \cdot \hat{Thr}_{ijk}^{(t)}=B_{ij}^{max}, \label{eq:constr3} \\ 
&\sum_{j=1}^{N} \sum_{i=1}^{|G_{j}|} x_{ijk}^{(t)} \cdot \lceil \frac{r_{ijk}^{(\lceil t/C \rceil)}}{Thr_{ijk}^{(t)}} \rceil \cdot W_{k}^{(t)} \leq W_{k}^{(t)}, \nonumber \\
& \hspace{30mm} \forall 1 \leq k \leq K, \hspace{3mm} A_{ij} \leq t \leq D_{ij} \label{eq:constr4}  
\end{align}
\begin{align}
& 0  < B_{ij}^{(t)}  \leq B_{ij}^{max}, \hspace{5mm} \forall A_{ij} \leq t \leq D_{ij}
\label{eq:constr5}  \\ 
& r_{ijk}^{(p)} \in Q_{k}, \hspace{2mm} x_{ijk}^{(t)} \in \{0,1\}, \hspace{8mm} \forall 1 \leq k \leq K,   \nonumber \\ 
& \hspace{20mm} A_{ij} \leq t \leq D_{ij}, 1 \leq p \leq |D_{ij}-A_{ij}|/C \label{eq:constr6}   
\end{align} 

In \eqref{eq:opt}, we obtain the average video quality over $|D_{ij}-A_{ij}|/C$ downloaded chunks by client $i \in G_{j}$ with \eqref{eq:1}.
\begin{align}
\label{eq:1}
AQ_{ij}=\frac{C}{|D_{ij}-A_{ij}|}\sum_{p=1}^{|D_{ij}-A_{ij}|/C} \sum_{k=1}^{K} x_{ijk}^{(A_{ij}+(p-1) \cdot C)} \cdot q_{ijk}^{(p)}
\end{align}

We denote the startup delay by $Delay_{ij}$ which is the time delay to reach the target buffer filling level $B_{ij}^{max}$ (in Kb) for client $i \in G_{j}$. Consequently, we have the constraint \eqref{eq:constr3}, where $\hat{Thr}_{ijk}^{(t)}$ is the effective data throughput (in Kbps) received by client $i \in G_{j}$ from server $k$ at time slot $t$. Denoted by $Thr_{ijk}^{(t)}$ as the theoretical throughput over the wireless link, we employ the simple path attenuation model $Thr_{ijk}^{(t)} \propto P_{max}/d_{ijk}^{(t) \alpha}$ for its computation. 
$P_{max}$ is the maximum transmission power of the base station, $d_{ijk}^{(t)}$ denotes the physical distance between the client $i \in G_{j}$ and base station $k$ at time slot $t$ and $\alpha$ is the path loss exponent parameter which is normally between 2 and 5. We note that the effective share throughput of client $i$ is computed by the relation $\hat{Thr}_{ijk}^{(t)}=Thr_{ijk}^{(t) 2}/\sum_{\forall i^{\prime} \in G_{j^{\prime}}} x_{i^{\prime}j^{\prime}k}^{(t)} \cdot Thr_{i^{\prime}j^{\prime}k}^{(t)}$ where the summation in denominator is taken over all clients $i^{\prime} \in G_{j^{\prime}}$ which have been assigned to base station $k$ at time slot $t$.   

The accumulated quality switching for client $i \in G_{j}$ during the streaming session is obtained with \eqref{eq:5}.
\begin{align}
\label{eq:5}
E_{ij}=\sum_{p=2}^{(D_{ij}-A_{ij})/C} \sum_{k=1}^{K} \{&x_{ijk}^{(A_{ij}+(p-1) \cdot C)} \cdot q_{ijk}^{(p)} \nonumber  \\
&-x_{ijk}^{(A_{ij}+(p-2) \cdot C)} \cdot q_{ijk}^{(p-1)}\}
\end{align}     

As for avoiding stall events, we assume that the player starts to play the video after the startup phase. Given $\hat{Thr}_{ijk}^{(t)}$, 
the buffer level (in Kb) of client $i \in G_{j}$ at time slot $t$ is given by \eqref{eq:3}.  
\begin{align}
\label{eq:3}
B_{ij}^{(t)}=
\begin{cases}
 B_{ij}^{(t-1)}+\hat{Thr}_{ijk}^{(t)}  , \hspace{6mm} A_{ij} \leq t \leq A_{ij}+Delay_{ij}  \\
 B_{ij}^{(t-1)}+\hat{Thr}_{ijk}^{(t)}-r_{ijk}^{(p)} , \hspace{2mm} A_{ij}+Delay_{ij} < t \leq D_{ij}   
 \end{cases}
\end{align}
In \eqref{eq:3}, $r_{ijk}^{(p)}$ is the allocated bitrate for the currently played out chunk with index $p$. Accounting for the arrival time of client and initial playback delay, the index $p$ of the chunk played out at time slot $t > A_{ij}+Delay_{ij}$ is equal to $p=\lceil (t-A_{ij}-Delay_{ij})/C \rceil$. Thus, we obtain \eqref{eq:constr5} as the constraint for buffer occupancy, which simply states that for client $i \in G_{j}$ it must be non-negative and also kept below or at the target filling level and ensures that no stall events happen provided that sufficient resources to sustain smallest available video bitrates for all clients exist. 

Proportional fair (PF) bitrate allocation for client $i \in G_{j}$ during a streaming session is defined in \eqref{eq:6}.   
\begin{align}
\label{eq:6}
PF_{ij}=log (\sum_{t=A_{ij}}^{D_{ij}} \sum_{k=1}^{K} x_{ijk}^{(t)} \cdot r_{ijk}^{(\lceil (t-A_{ij})/C \rceil)}) 
\end{align} 

\begin{algorithm}[t]
\small
\caption{GreedyMSMC}
\label{alg1}
\begin{algorithmic}[1]
\small 
\Require \parbox[t]{\dimexpr\linewidth-\algorithmicindent}{$|T|, N, K, Q_{k}:$ Number of scheduling time slots, number \\
of groups and DASH servers, set of available discrete bitrates on server $1 \leq k \leq K$. \strut} 
\Ensure \parbox[t]{\dimexpr\linewidth-\algorithmicindent}{Binary allocation $x_{ijk}^{(t)}$ and integer bitrate allocation $r_{ijk}^{(t)}$ for each client $i \in G_{j}, 1 \leq j \leq N$, server $1 \leq k \leq K$ and time slot $1 \leq t \leq |T|$, $totalUtility$\strut}
\vspace{0.06em}
\State \parbox[t]{\dimexpr\linewidth-\algorithmicindent} {$totalUtility=0$ \strut} 
\For{each time slot $1 \leq t \leq |T|$}
 \For{each group $1 \leq j \leq N$}
   \For{each client $i \in G_{j}$ such that $A_{ij} \leq t \leq D_{ij}$}
      \State \parbox[t]{\dimexpr\linewidth-\algorithmicindent}{$Utility_{ij}=-\infty$; \strut}    
      \State{**\emph{in the middle of chunk}**}
      \If{$(t-A_{ij})$ \textbf{mod} $C \neq 1$}
        \State \parbox[t]{\dimexpr\linewidth-\algorithmicindent}{\textbf{Select} the same bitrate and server \\
         \indent \hspace{8mm} as for time slot $t-1$; \strut}
         \State{\textbf{Update} $Buffer.Level$ and $Delay$;}
      \EndIf
      \State{**\emph{at the beginning of chunk}**}
      \If{$(t-A_{ij})$ mod $C == 1$}
         \If{$Buffer.Filling == False$}
             \State \parbox[t]{\dimexpr\linewidth-\algorithmicindent}{\textbf{Run} Subroutine Startup Phase; \strut}
         \EndIf
         \If{$Buffer.Filling == True$}
             \State \parbox[t]{\dimexpr\linewidth-\algorithmicindent}{\textbf{Run} Subroutine Steady State; \strut}
         \EndIf 
      \EndIf
  \If{$t == D_{ij}$}
      \State  \parbox[t]{\dimexpr\linewidth-\algorithmicindent}{$totalUtility=totalUtility+Utility_{ij}$ \strut}   
  \EndIf   
 \EndFor
 \EndFor
 \EndFor
\State  \parbox[t]{\dimexpr\linewidth-\algorithmicindent}{\textbf{Return} $totalUtility$; \strut}
\end{algorithmic}
\end{algorithm}
The maximization of $PF_{ij}$ is subject to the available resources at the base stations at each time slot, i.e. constraint \eqref{eq:constr4}. 

We compute the standard deviation of resource utilization on $K$ servers from the average utilization as a criteria for measuring the load balancing on servers. Denoted by $U_{ijk}$, the ratio of occupied resources by client $i \in G_{j}$ to the total available resources on the bottleneck link of base station $1 \leq k \leq K$: 
\begin{align}
U_{ijk}=\sum_{t=A_{ij}}^{D_{ij}} (1/W_{k}^{(t)})(x_{ijk}^{(t)} \cdot \lceil \frac{r_{ijk}^{(\lceil (t-A_{ij})/C \rceil)}}{Thr_{ijk}^{(t)}} \rceil \cdot W_{k}^{(t)}) 
\end{align}

Note that the allocated bitrate for the client is divided by its actual physical (theoretical) transmission rate to be converted to the amount of resources which are allocated by base station to the client. Let $\bar{U_{ij}}=(1/K)\sum_{k=1}^{K} U_{ijk}$ denote the average utilization efficiency of client $i \in G_{j}$ on $K$ bottleneck links. For each client $i \in G_{j}$, the objective of load balancing is to decide on servers and the allocate bitrates to the client depending on its physical location and channel quality in such a way that the standard deviation of loads incurred by the client i.e. $SD_{ij}=\sqrt{(1/K)\sum_{k=1}^{K} (U_{ijk}-\bar{U_{ij}})^2}$ is minimized.      

As for the remaining constraints, \eqref{eq:constr1} states that at any time instant $t$, the DASH client is allocated to only one server for downloading its current video chunk and \eqref{eq:constr2} enforces that the client receives one complete chunk of video upon its access to the allocated server. Finally, constraint \eqref{eq:constr6} specifies that the discrete allocated bitrate for a requested chunk of video which is resided on a given server belongs to the set of bitrate 
\begin{tabular}{p{8.25 cm}}
\hline 
\textbf{Subroutine 1:} Startup Phase \\
\hline \\
\end{tabular}
\begin{algorithmic}
\small 
\vspace{-1em}
     \For{each video server $1 \leq k \leq K$ from the sorted set}
           \For{each $r \in Q_{k}$}
                \State{**\emph{feasible resource allocation}**}
                \State{\textbf{if} the allocation of bitrate $r$ is feasible}
                \State \parbox[t]{\dimexpr\linewidth-\algorithmicindent} {\indent \hspace{1mm} \textbf{Compute} $AQ_{ij}, E_{ij}$ from (8) and (9); \strut}
                \State \parbox[t]{\dimexpr\linewidth-\algorithmicindent} {\indent \hspace{1mm} \textbf{Compute} $PF_{ij}$ from (11) and $SD_{ij}$; \strut}
                \State \parbox[t]{\dimexpr\linewidth-\algorithmicindent} {\indent \hspace{1mm} $maxUtility=$ Objective value of function (1); \strut} 
                \State \parbox[t]{\dimexpr\linewidth-\algorithmicindent}{\indent \hspace{2mm} \textbf{if} $maxUtility > Utility_{ij}$}
                \State \parbox[t]{\dimexpr\linewidth-\algorithmicindent}{\indent \hspace{5mm} $Utility_{ij}=maxUtility;$ \strut}
                \State \parbox[t]{\dimexpr\linewidth-\algorithmicindent}{\indent \hspace{5mm} $selectedRate=r; 
                \hspace{3mm} selectedServer=k;$ \strut}
         \EndFor
        \EndFor
       \State{\textbf{Allocate} $selectedServer$ and $selectedRate$ to client $i \in G_{j}$}
       \State{\textbf{Update} $Buffer.Level$, $Buffer.Filling$ and $Delay$;}
\end{algorithmic}
\begin{tabular}{p{8.25 cm}}
\hline \\
\end{tabular}
resolutions offered by that server and also the binary allocation of the client to a server at each time slot.

\section{Centralized Scheduling Algorithm}
\label{sec:solution}

The joint optimization problem formulated in \eqref{eq:opt}-\eqref{eq:constr6} belongs to the class of NP-hard problems because it contains integer decision variables. 
Although the brute-force strategy can be applied for the offline case when all the clients' information are available in advance, 
however, with a total of $S$ active clients at each time slot and $K$ servers, the computational complexity of this approach is $O(\frac{|T|}{C} \cdot K^S \cdot Q)$. Here, $|T|$ is the number of time slots, $C$, the chunk size and $Q$ is the number of different video bitrates available. That means the complexity of exhaustive approach grows dramatically with the increase in the number of servers or clients making it impractical for large scale deployments. Therefore, we devise an efficient online and centralized greedy algorithm, which we name \textbf{GreedyMSMC} with the pseudocode in Algorithm \ref{alg1}. It is noteworthy to mention that the high computational complexity of the proposed centralized algorithm specially in large scale deployments, can be degraded using its decentralized implementation. It should be also mentioned that although the decentralized algorithm improves the computation time but, it sacrifices slightly the performance gain.    

With dynamic arrival and departure of clients, for each client which is active in the current slot, the algorithm first sorts the set of available base stations based on their closeness to the client's location. It then checks for the possibility of allocating the client to each base station and selects in a greedy manner the target base station and a sustainable bitrate from its associated server such that the constraints \eqref{eq:constr2}-\eqref{eq:constr6} are satisfied and the objective function \eqref{eq:opt} achieves locally the maximum utility. We note that in the selection of bitrate for the current chunk, the algorithm takes into account the instantaneous client's buffer occupancy in order to avoid the possibility of happening a stalling event. 
In the startup phase, Subroutine 1 is run in order to quickly fill up the buffer and then, the algorithm runs the steady phase shown in Subroutine 2 in which both average quality and bitrate switching are accounted for when selects bitrates. 
With $S$ active clients and $K$ servers, the worst case complexity of the greedy algorithm is $O(\frac{|T|}{C} \cdot S \cdot K \cdot Q)$, which is a significant reduction in complexity compared to the exhaustive search.  
\begin{tabular}{p{8.25 cm}}
\hline 
\textbf{Subroutine 2:} Steady State \\
\hline \\
\end{tabular}
\begin{algorithmic}
\small    
      \State{$rate=$ Bitrate of current streaming chunk;}
      \For{each video server $1 \leq k \leq K$ from the sorted set}
        \State \parbox[t]{\dimexpr\linewidth-\algorithmicindent}{\textbf{if} 
         $0 < Buffer.Level-rate \leq B_{ij}^{max}$ \textbf{then} \strut}
         \For{each bitrate $r \in Q_{k}$}
            \State  \parbox[t]{\dimexpr\linewidth-\algorithmicindent}{\textbf{Execute} same codes lines (4)-(10) in Subroutine  \\
            \indent \hspace{7mm} Startup Phase using objective function (1); \strut}
         \EndFor
      \EndFor
     \State{\textbf{Allocate} $selectedServer$ and $selectedRate$ to clinet $i \in G_{j}$;}
     \State{\textbf{Update} $Buffer.Level$}
\end{algorithmic}
\begin{tabular}{p{8.25 cm}}
\hline \\
\end{tabular}

\section{Evaluation}
\label{sec:eval}

In this section, we evaluate the performance of GreedyMSMC through simulations. 
We compared the results of GreedyMSMC to those obtained with two client-based adaptation heuristics, namely buffer-based adaptation (BBA) \cite{Huang2014} and rate-based adaptation (RBA) \cite{Mangla2016}. The simulator and all the algorithms are implemented on MATLAB. 

\subsection{Simulation Setup}

We consider the scheduling of DASH clients during one hour with time duration $|T|=3600$ $Seconds$. For the network setup, we assume a rectangular area with size $400m \times 1km$ where base stations are located with equal distances and the clients are randomly distributed around the base stations. Clients arrival time is uniformly distributed within the first $20$ $min$ and they depart after an active session which its length duration is chosen from the uniform distribution $[1000 Seconds, 2000 Seconds]$. All clients are stationary. The video is divided into $C=5$ $sec$ chunks. Each video chunk is available in six different bitrates $[60$ $kbps,$ $90$ $kbps,$ $110$ $kbps,$ $130$ $kbps,$ $170$ $kbps,$ $220$ $kbps]$ with the same replication on each edge server. In the simulations, we consider 12 number of servers unless otherwise stated, and, the number of clients vary from 100 to 500. The maximum transmission power of each base station associated with a video server is fixed at 3.6 $\times$ $10^6 mW$ and the path loss exponent $\alpha=2$ is considered in the path attenuation model. With time slot duration $1$ $Second$ and total available per slot bandwidth $U[90 KHz, 180 KHz]$, the total LTE resource blocks per slot at each base station follows the uniform distribution $U[100,200]$ \cite{Sesia2009}. The tuning parameters in the objective function in \eqref{eq:opt} are set to $\beta=0.7$, $\theta=0.25$, $\mu=0.05$ and $\rho=0.7$, $\omega=0.05$ and $\gamma=0.25$ in the simulations. It is noteworthy to mention that these values have been chosen properly after our simulations to study the impact of varying the tuning parameters on the performance gain, although, we have not reported those results here due to the space limitations. We compare our network-assisted method to the two following client-based adaptation strategies which both of them assign each client to the closest neighborhood base station for the whole video streaming session of the client.  

\textbf{Buffer Based Adaptation (BBA)} \cite{Huang2014} means that each client independently selects the bitrate for the next chunk to download based on instantaneous buffer occupancy level, i.e, the amount of video data in the playback buffer of the client 
\begin{figure}[t] 
  \begin{subfigure}[b]{0.47\linewidth}
    \centering
     \includegraphics[width=1\linewidth]{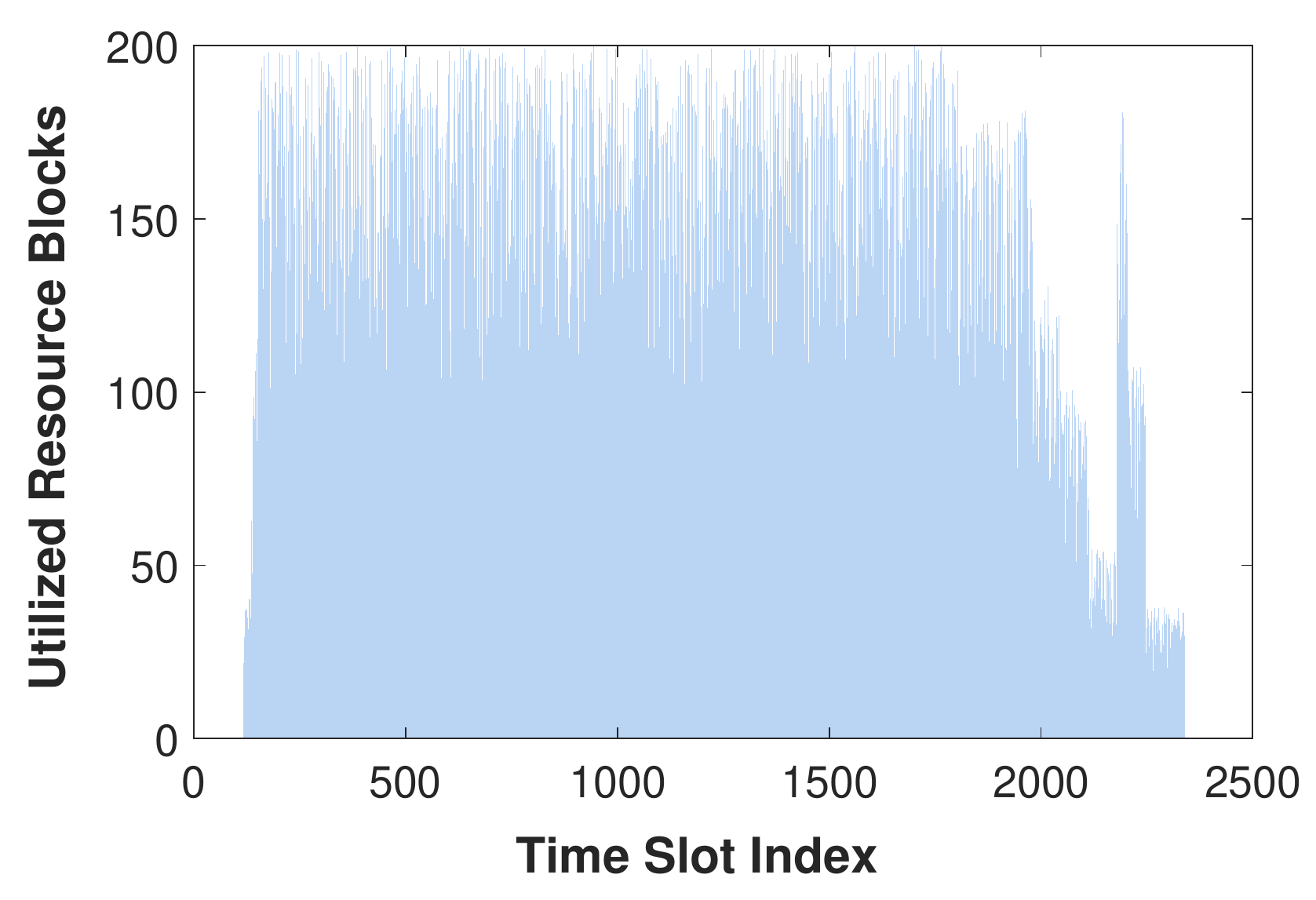}  
    \caption{Pattern of utilized resources} 
    \label{fig:utilizedresources}
    \vspace{2ex}
  \end{subfigure}
  \begin{subfigure}[b]{0.47\linewidth}
    \centering
   \includegraphics[width=1\linewidth]{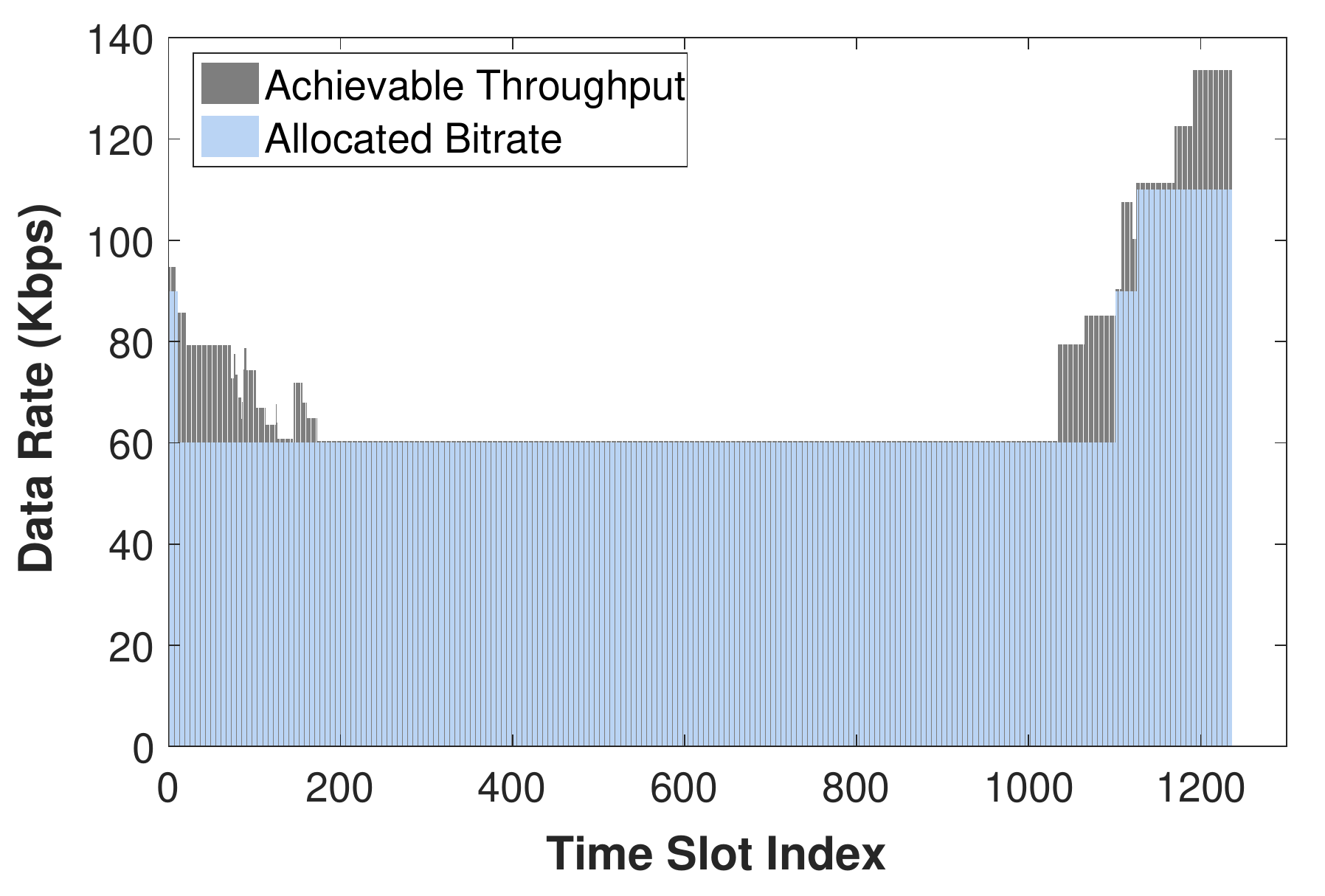} 
    \caption{Achievable throughput and allocated bitrate} 
    \label{fig:throughputbitrate}
    \vspace{0.2ex}
  \end{subfigure} 
  \caption{(a) The pattern of utilized resource blocks of a base station (b) Achievable throughput/bitrate for a client.} 
  \label{fig:ResourceUtilization}
\end{figure}
at each time slot. 
The heuristic allocates the highest bitrate for the first chunk and then looks at the current client's buffer level to decide on the bitrate for the next video chunk to be downloaded. The heuristic considers five different thresholds which are equal to $2/6$, $3/6$, $4/6$, $4.5/6$ and $5/6$ fraction of the maximum buffer filling level and depending on the buffer level, it chooses the most closest bitrate from the server. 

\textbf{Rate Based Adaptation (RBA)} \cite{Mangla2016} works so that each client makes chooses the highest sustainable bitrate among the available ones based on throughput obtained when downloading the previous $m$ chunks. In particular, RBA computes a moving average of the download rate of the last consecutive $m$ chunks to determine the bitrate for the next video chunk to be downloaded. The bitrate for chunk $i > m$ is obtained using the moving average $(1/m)\sum_{j=i-m}^{i-1} r_{j}$.

\subsection{Resource Utilization and Bitrate Allocation}

As the first result, we have shown in Fig. \ref{fig:utilizedresources} the pattern of utilized resource blocks of one randomly chosen base station during 2500 time slots using GreedyMSMC algorithm. The amount of utilized resource blocks of the base station at each time slot can vary depending on the instantaneous number of allocated clients. As we can see from the pattern, the utilization level can reach up to 200 which is the maximum number of available resource blocks on the base station at each time slot.

We have also shown in Fig. \ref{fig:throughputbitrate}, the pattern of achievable throughput and the allocated bitrates using GreedyMSMC algorithm for one randomly chosen client during 1300 time slots. As we can see, the throughput of the client is less for the intermediate time slots where most of the clients are active within their streaming session. 
We can also see that the proposed algorithm behaves well in determining the best sustainable bitrate from the discrete set by observing the effective obtainable throughput at each time slot.

\subsection{Comparison to Client-based Adaptation Approaches}

In this section, we compare GreedyMSMC algorithm with two client-based DASH heuristics BBA and RBA in terms of the average achievable throughput for clients and the deviation of resource utilization 
\begin{figure}[t] 
  \begin{subfigure}[b]{0.5\linewidth}
    \centering
     \includegraphics[width=1\linewidth]{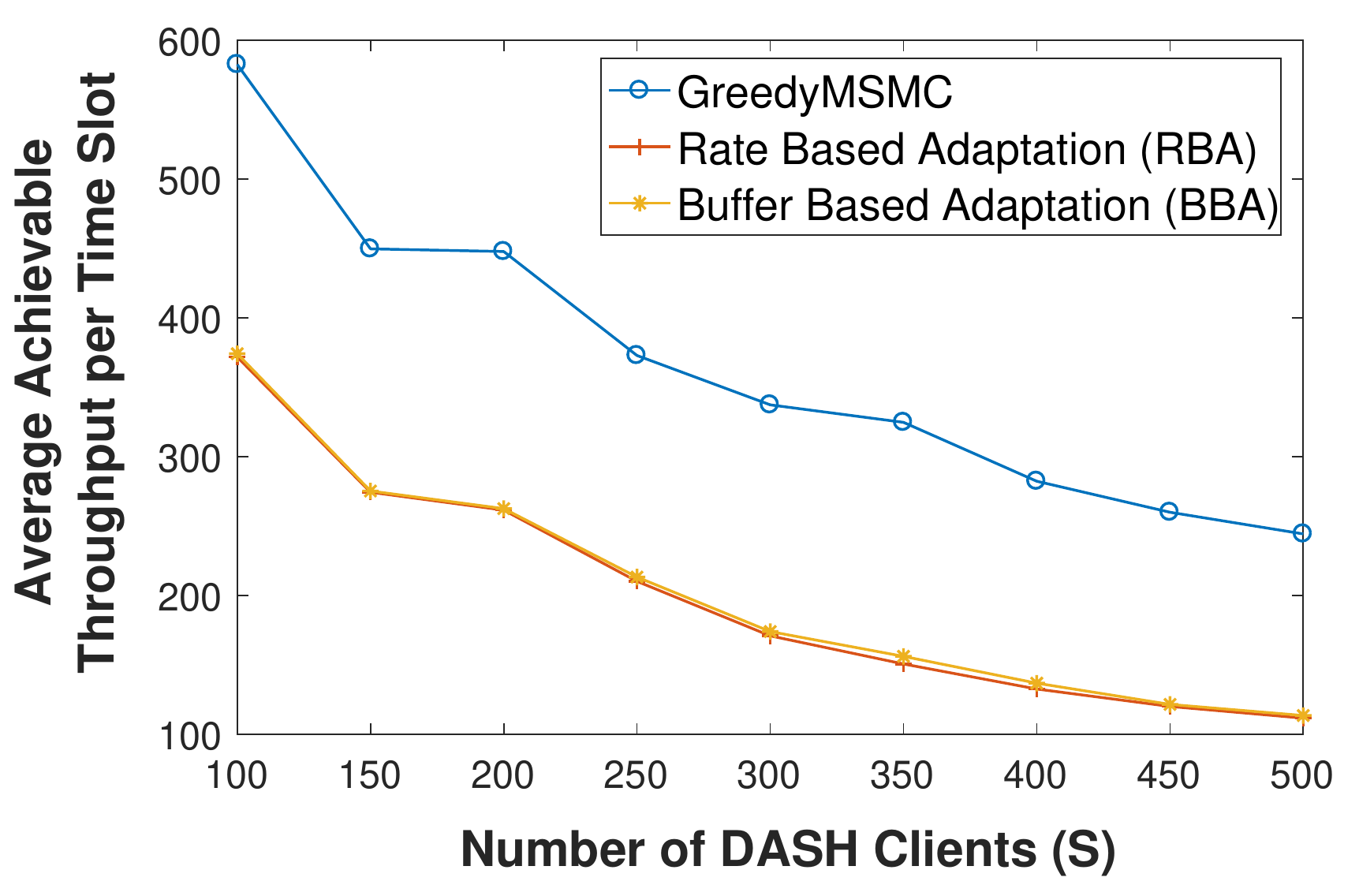}  
    \caption{Average achievable throughput} 
    \label{fig:AchievableThroughput}
    \vspace{2ex}
  \end{subfigure}
  \begin{subfigure}[b]{0.5\linewidth}
    \centering
   \includegraphics[width=1\linewidth]{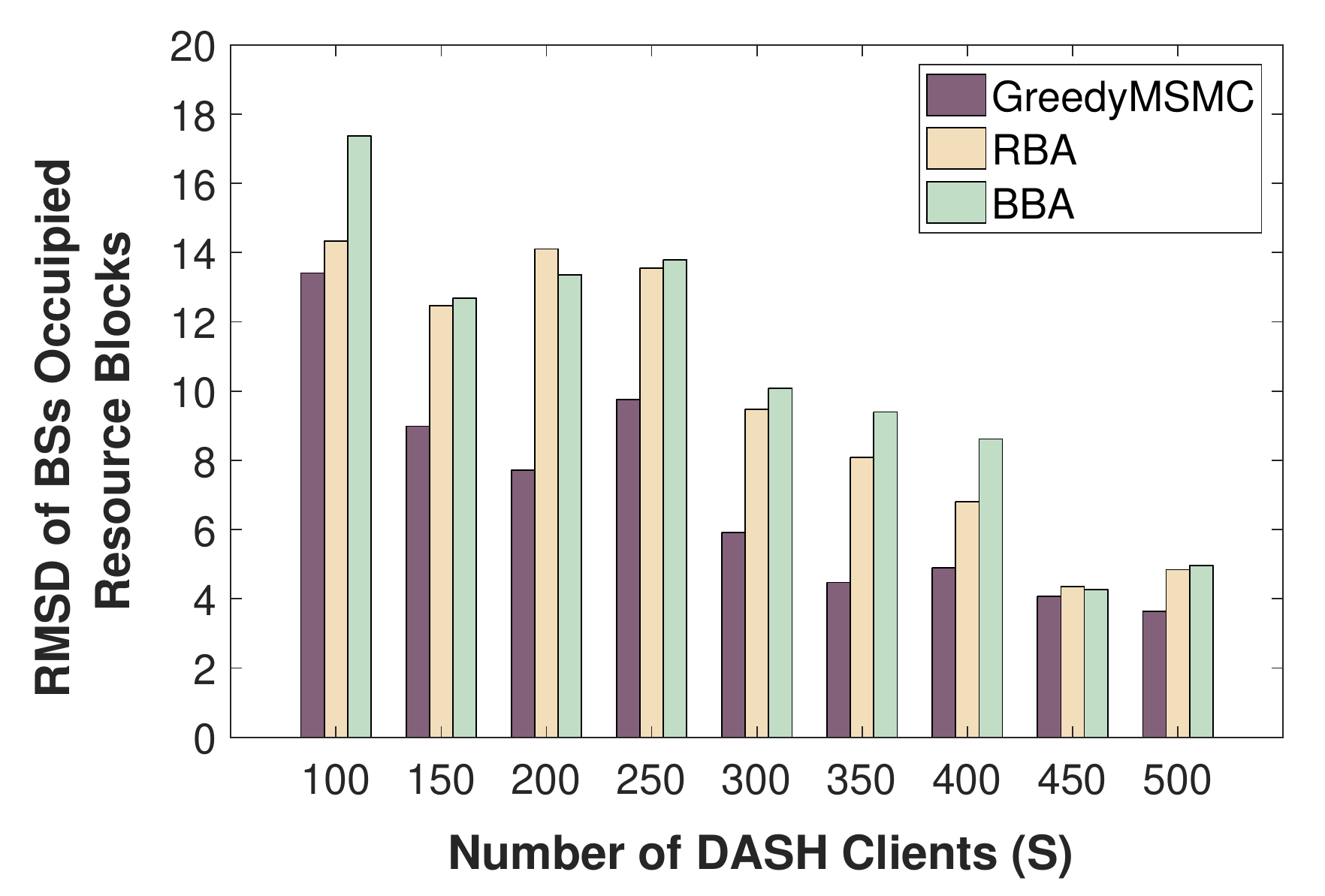} 
    \caption{Resource utilization deviation} 
    \label{fig:UtilizationDeviation}
    \vspace{2ex}
  \end{subfigure} 
  \caption{Comparison between GreedyMSMC and two client-based DASH heuristics in term of the achievable throughput per time slot and the resource utilization deviation (rmsd).}
  \label{fig:compariosnOPT}
\end{figure}
among the base stations. Note that in the implementation of RBA heuristic, we set $m=3$ as the number of previously observed chunks when estimating the achievable throughput for each current chunk. 

Fig. \ref{fig:AchievableThroughput} shows the comparison between GreedyMSMC algorithm and two client-based adaptation heuristics in term of the average achievable throughput per time slot for different number of DASH clients. As we can see, the clients achieve significantly higher effective throughput using the proposed algorithm compared to the purely client-based heuristics. The reason is that allocating each client merely to the closest base station during the whole active session of the client will result in lowering the average throughput especially under the high dynamic arrival and departure of clients. In contrast, GreedyMSMC algorithm takes into account the current load of the base stations and seeks for the most suitable base station for the client where the higher throughput can be obtained. 
It is also seen that the average throughput drops as the number of clients increases which is due to high competition for sharing the available resources of base stations.         

Fig. \ref{fig:UtilizationDeviation} shows the comparison in term of the deviation of the utilized resource blocks among the base stations. To measure the utilization efficiency, we employ the root mean square deviation (RMSD) of the utilized resources among the base stations during the whole streaming session of clients. As we can see from the result, using the proposed algorithm results in lesser utilization deviation since GreedyMSMC algorithm allocates clients to appropriate base stations in order to minimize the utilization deviation.  

\begin{figure}[t] 
  \centering
  \begin{subfigure}[b]{0.48\linewidth}
    \centering
     \includegraphics[width=1\linewidth]{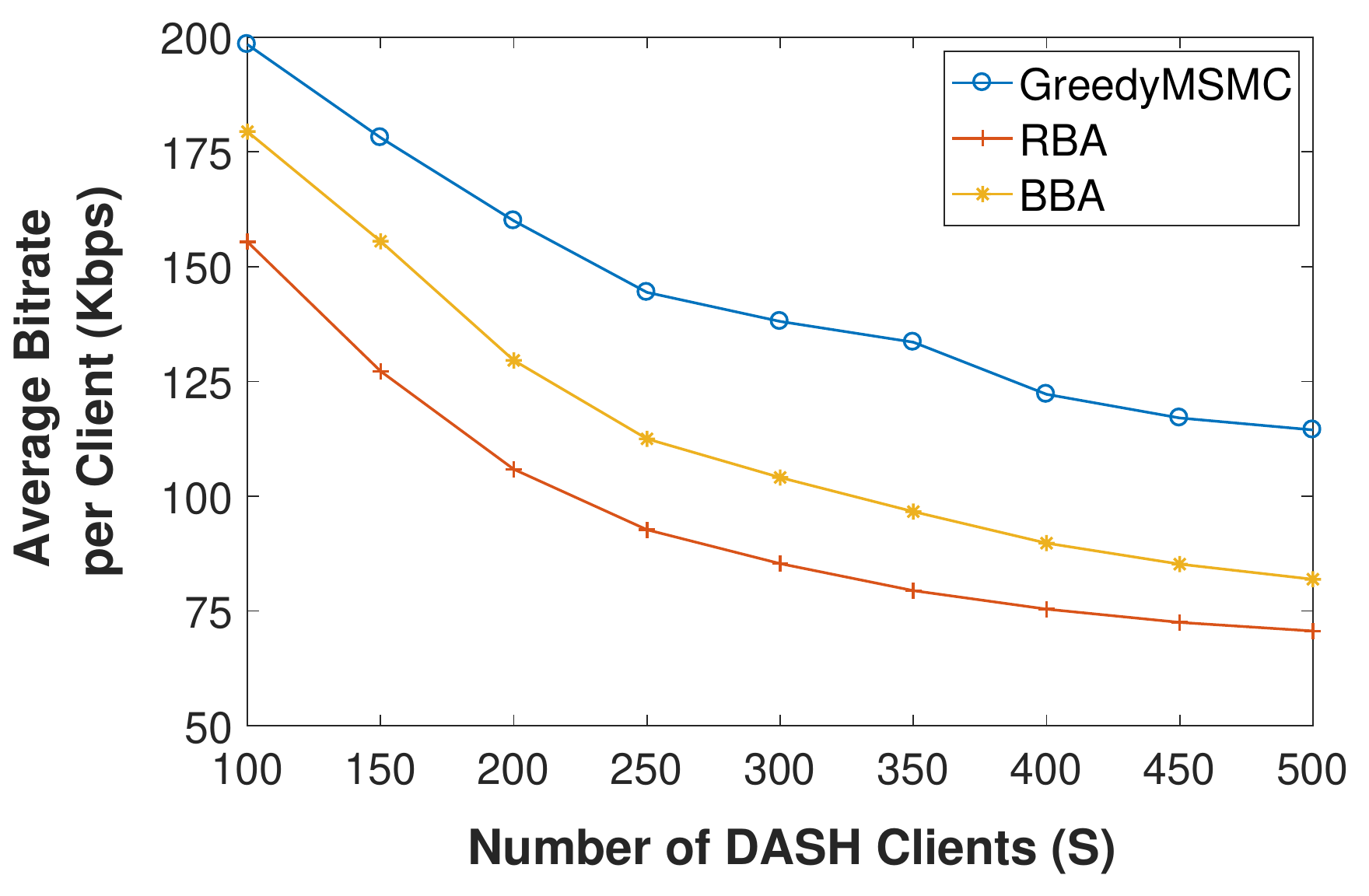}  
    \caption{Average video bitrates} 
    \label{fig:bitrate}
    \vspace{0.5ex}
  \end{subfigure}
  \begin{subfigure}[b]{0.48\linewidth}
    \centering
     \includegraphics[width=1\linewidth]{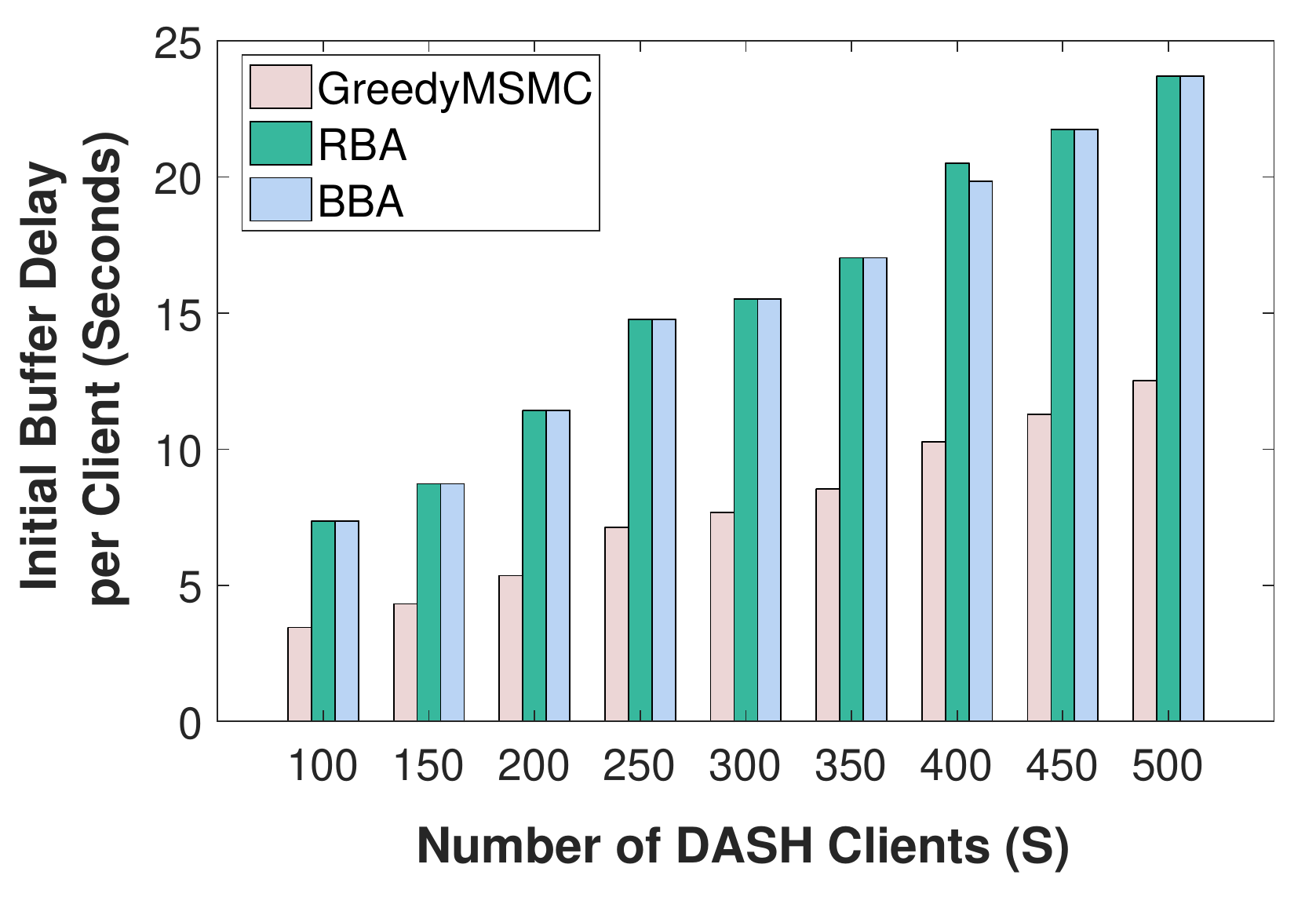}  
    \caption{Initial buffer delay}
    \label{fig:delay}
  \end{subfigure}
  \begin{subfigure}[b]{0.48\linewidth}
    \centering
   \includegraphics[width=1\linewidth]{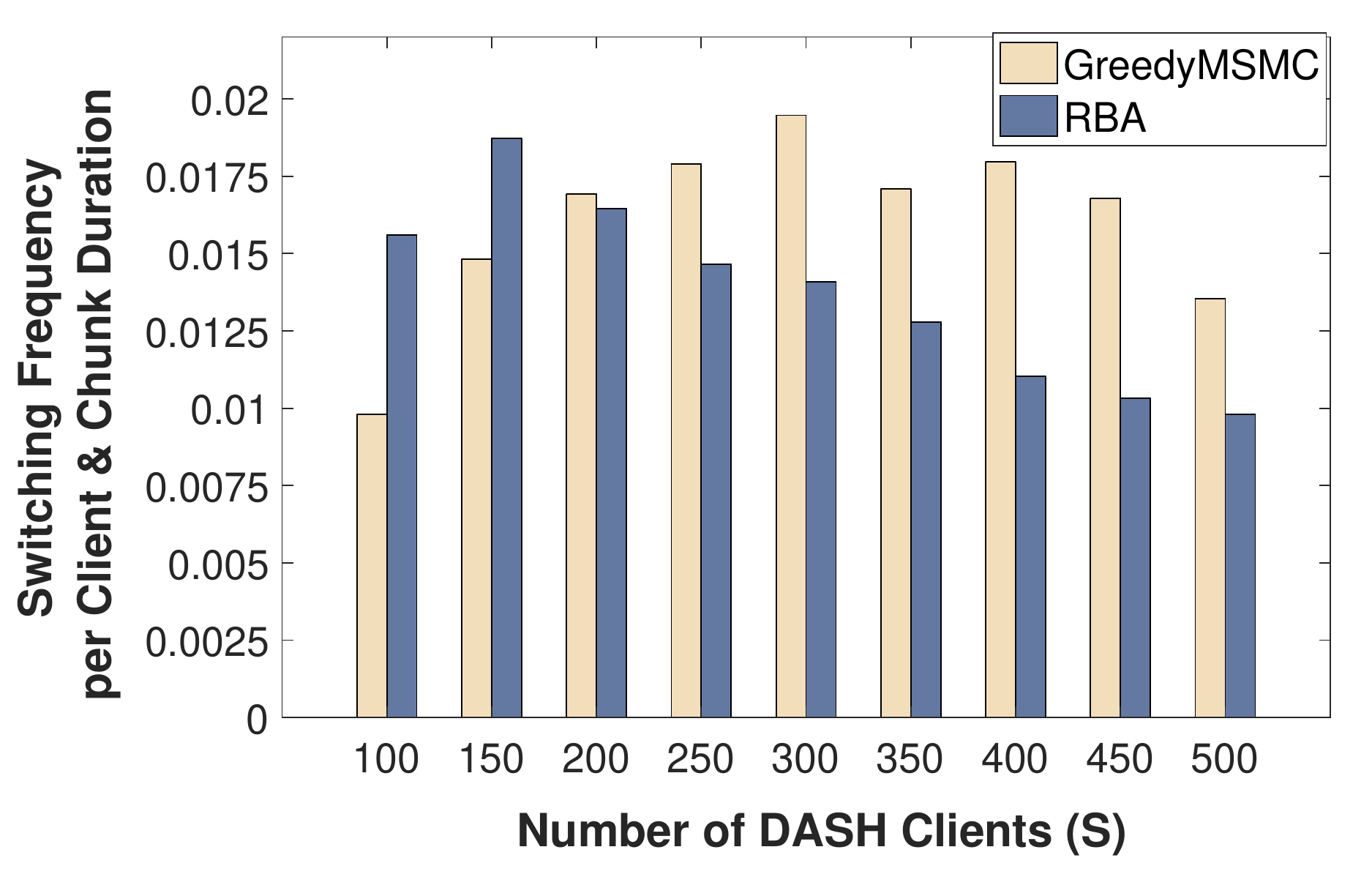} 
    \caption{Bitrate switching frequency} 
    \label{fig:switchingfrequency}
    \vspace{0.4ex}
  \end{subfigure}
  \begin{subfigure}[b]{0.48\linewidth}
    \centering
   \includegraphics[width=1\linewidth]{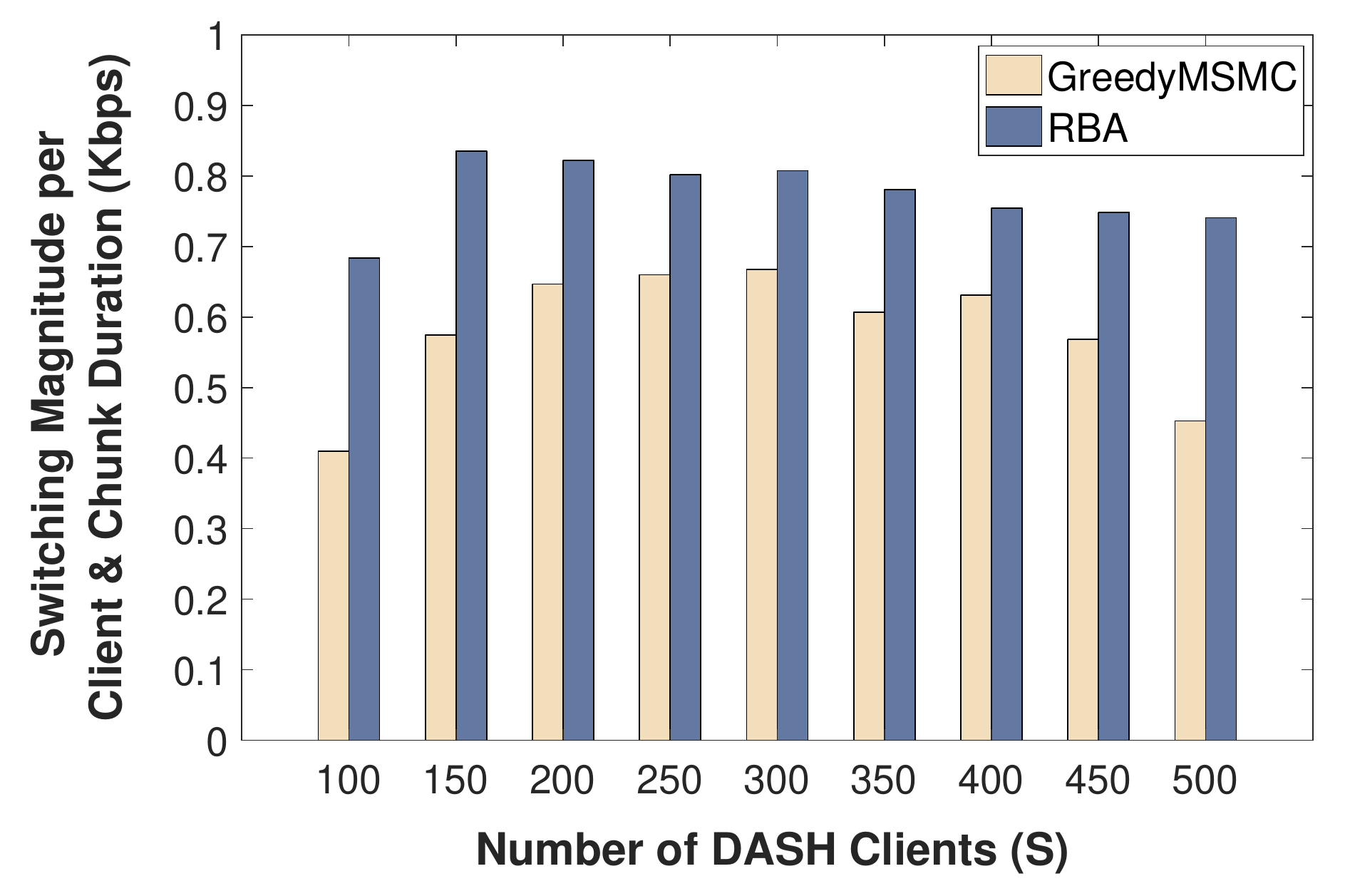} 
    \caption{Bitrate switching magnitude} 
    \label{fig:switchingmagnitude}
  \end{subfigure} 
  \caption{The comparison between GreedyMSMC and two client-based DASH heuristics in terms of QoE metrics.}
  \label{fig:comparisonQoE}
\end{figure}

With the same dataset as the previous part, we have also compared GreedyMSMC algorithm with BBA and RBA in terms of the QoE metrics. Figure \ref{fig:comparisonQoE} shows that GreedyMSMC outperforms both heuristics in terms of average video bitrates and the initial buffer delay per client as well as the magnitude and frequency
of bitrate switching per chunk duration. As observed from Fig. \ref{fig:bitrate}, the improvement in average bitrate using the proposed algorithm is due to achieving the higher share of throughput for clients by taking into account the instantaneous load of base stations. 
Since in the startup phase, GreedyMSMC chooses the bitrates which minimize the gap between the instantaneous and the maximum buffer level, therefore, about 50\% reduction in initial buffer delay per client is achieved compared to client-based heuristics (Fig. \ref{fig:delay}).  

 
Fig. \ref{fig:switchingfrequency} and \ref{fig:switchingmagnitude} show respectively the frequency and the magnitude (Kbps) of bitrate switching per chunk duration during the whole active sessions of all clients. We have excluded from the charts the switching values for BBA which were around 10 times bigger than RBA in both frequency and magnitude. As an example to interpret the meaning of values on y-axis in Fig. \ref{fig:switchingfrequency}, for 100 number of clients and using RBA, the switching happens for around 1.6 percentage of all chunks and the magnitude of each switching will be around $0.7kbps$ according to Fig. \ref{fig:switchingmagnitude}. GreedyMSMC and RBA are both effective in significantly reducing the frequency and magnitude of bitrate switching compared to BBA. The reason is that the buffer occupancy level can highly fluctuate especially under high dynamic arrival and departure of clients which results in larger number of bitrate switching per client. It is also observed that although RBA exhibits less switching frequency for larger number of clients, however, it has bigger switching magnitude per chunk as we can see from Fig. \ref{fig:switchingmagnitude}. We should also acknowledge that the authors in \cite{Huang2014} have proposed a variation of BBA which can reduce to some extent the bitrate switching by having an estimation of the throughput variation for the future chunks.
 
\begin{figure}[t] 
    \centering
     \includegraphics[width=4.75cm]{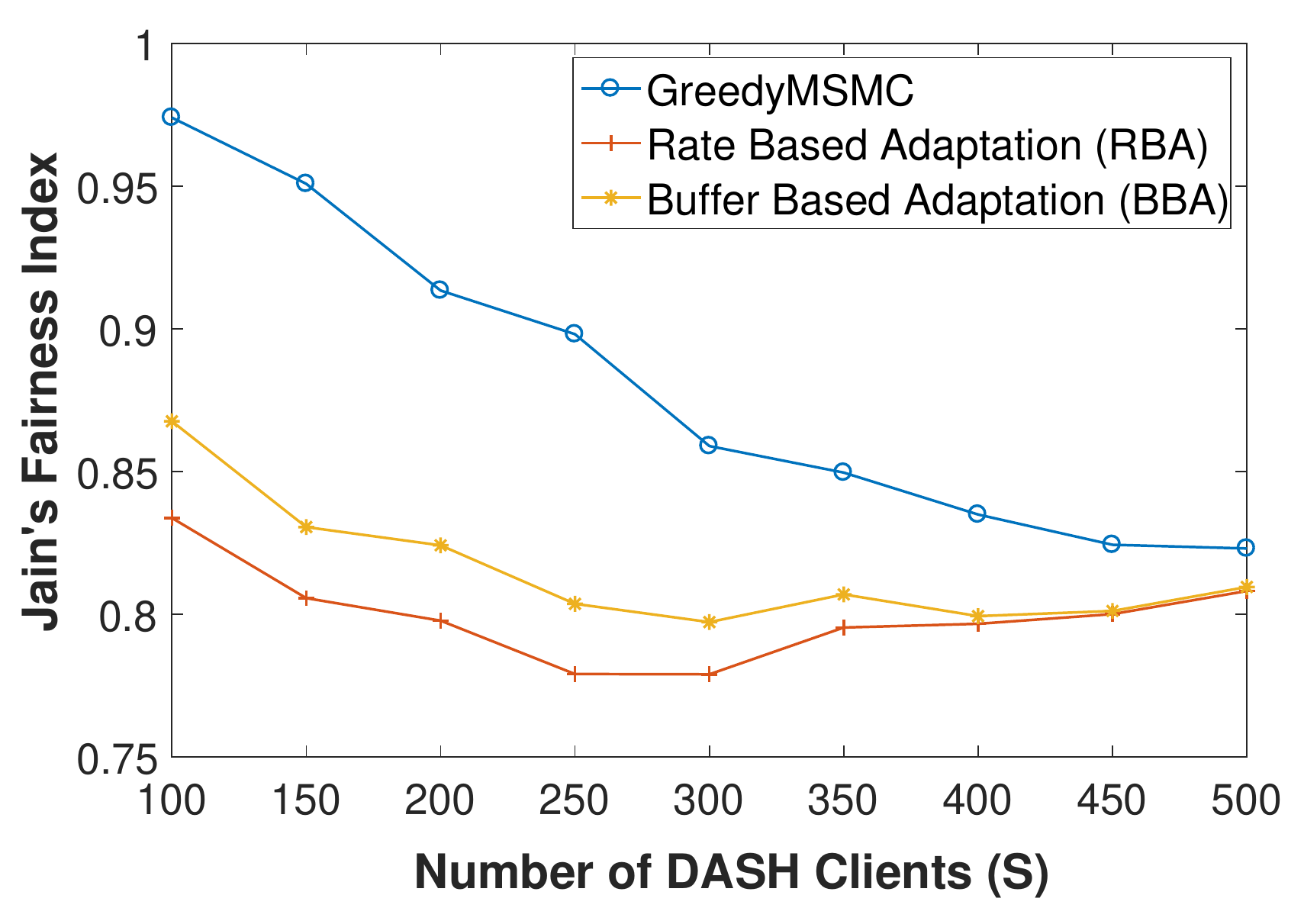}  
    \caption{Comparison in term of Jain's fairness index.} 
    \label{fig:fairnessindex}
\end{figure}

In Fig. \ref{fig:fairnessindex}, we have compared three adaptation approaches in term of fairness in the average bitrate that each client perceives during its active session. We employ the Jain's fairness index \cite{Bouten2014} which is defined as $JF=(\sum_{i} \bar{r}_{i})^2/(S \cdot \sum_{i} \bar{r}_{i}^2),$ where $S$ is the total number of clients and $\bar{r}_{i}$ denotes the average bitrate of client $1 \leq i \leq S$ during its streaming session. As an example, we have considered a scenario in which 30\% of clients are located far from the base stations while the remaining 70\% of clients are closer to the base stations and therefore are prone to get higher average bitrates. As we can see from the result, using GreedyMSMC results in better fairness index for different number of clients. This is because in contrast to client-based heuristics, the proposed algorithm strives to improve the average bitrate of far-away clients by exploring among the best possible base stations to allocate them.    
It is also seen that the fairness value drops as the number of clients increases which is due to increasing the degree of competition among them for the shared bandwidth. 

Finally, it is noteworthy to mention that although the fairness values of three approaches are closer for large number of clients, using our algorithm results in higher average bitrate (Fig. \ref{fig:bitrate}) which also confirms the trade-off between the fairness and the achievable bitrate. 

\section{Conclusion and Future Works}

In this work, we studied the use of network assisted adaptive video streaming to mobile clients from mobile edge servers. 
We proposed an optimization model to jointly maximize the QoE of individual clients, enforce proportional fair video bitrate selection between the clients, and balance load among video servers. We then design an efficient centralized scheduling algorithm to tackle the large scale optimization problem. Our simulation based evaluation results suggest that network assistance indeed helps to achieve better QoE and fairness.

As future work, we intend to study the impact of clients mobility and varying the video chunks and buffer size on the performance of the proposed framework. We also plan to design a decentralized solution to the optimization problem and compare its performance to the centralized one.


\section*{Acknowledgment}

This work has been financially supported by the Academy of Finland (grant numbers 278207 and 297892), Tekes - the Finnish Funding Agency for Innovation, and the Nokia Center for Advanced Research.



%

\end{document}